\documentclass[preprint2]{aastex}

\shorttitle{\indent \def Outflows in plume-like structures}
\shortauthors{Tian et al.}

\begin{document}

\title{Observation of High-speed Outflow on Plume-like Structures of the Quiet Sun and Coronal Holes with SDO/AIA}

\author{Hui Tian\altaffilmark{1}, Scott W. McIntosh\altaffilmark{1}, Shadia Rifal Habbal\altaffilmark{2}, Jiansen He\altaffilmark{3}}
\altaffiltext{1}{High Altitude Observatory, National Center for Atmospheric Research, P.O. Box 3000, Boulder, CO 80307; htian@ucar.edu; mscott@ucar.edu}
\altaffiltext{2}{Institute for Astronomy, University of Hawaii, 2680 Woodlawn Drive, Honolulu, HI 96822; shadia@ifa.hawaii.edu}
\altaffiltext{3}{School of Earth and Space Sciences, Peking University, 100871 Beijing, China; jshept@gmail.com}

\begin{abstract}
Observations from the Atmospheric Imaging Assembly (AIA) onboard the Solar Dynamics Observatory (SDO) reveal ubiquitous episodic outflows (jets) with an average speed around 120 km s$^{-1}$ at temperatures often exceeding a million degree in plume-like structures, rooted in magnetized regions of the quiet solar atmosphere. These outflows are not restricted to the well-known plumes visible in polar coronal holes, but are also present in plume-like structures originating from equatorial coronal holes and quiet-Sun regions. Outflows are also visible in the ``inter-plume'' regions throughout the atmosphere. Furthermore, the structures traced out by these flows in both plume and inter-plume regions continually exhibit transverse (Alfv\'enic) motion. Our finding suggests that high-speed outflows originate mainly from the magnetic network of the quiet Sun and coronal holes, and that the plume flows observed are highlighted by the denser plasma contained therein. These outflows might be an efficient means to provide heated mass into the corona and serve as an important source of mass supply to the solar wind. We demonstrate that the quiet-Sun plume flows can sometimes significantly contaminate the spectroscopic observations of the adjacent coronal holes - greatly affecting the Doppler shifts observed, thus potentially impacting significant investigations of such regions.
\end{abstract}


\section{Introduction}
The ultraviolet emission formed in the polar region of the solar atmosphere is dominated by bright plume-like structures\citep[see a review by][]{Kohl2006}. Propagating intensity perturbations with a period of five to thirty minutes have been frequently identified in polar plumes \citep[e.g.,][]{Ofman1997,DeForest1998}. These outward propagating features have been widely interpreted as slow magneto-acoustic waves, since the speed is close the coronal sound speed of around 150~km s$^{-1}$ \citep[e.g.,][]{Ofman1999,Oshea2007,Banerjee2009,Gupta2010,KrishnaPrasad2011}.  Quasi-periodic intensity variations were also detected in inter-plume regions above the limb and again were interpreted as compressional waves \citep{Ofman2000,Banerjee2001}. Recently \cite{McIntosh2010} noted that these propagating features are similar in intensity enhancement, periodicity and velocity to the ubiquitous high-speed upflows inferred from spectroscopic observations in other magnetic regions of the solar atmosphere \citep{DePontieu2009,McIntosh2009a,McIntosh2009b,McIntosh2011,DePontieu2010,DePontieu2011,Tian2011a,Tian2011b,Martnez-Sykora2011}. They interpreted them as quasi-periodically driven high-speed outflows and suggested that they are a source of heated mass to the corona and fast solar wind, rooted in the lower atmosphere. There is also a suggestion that they are warps in two-dimensional sheet-like structures \citep{Judge2011}.

For a long time it was believed that plumes are polar features. However, \cite{Wang1995} and \cite{Woo1996} identified plumes in low-latitude coronal holes by using {\it Skylab} EUV observations and  {\it Ulysses} radio ranging measurements, respectively. They both concluded that plumes are a feature common to coronal holes at any latitude. 
Recently, \cite{Wang2008} performed a detailed study of  the morphology of low-latitude coronal-hole plumes and their close relationship with small bipoles in photospheric magnetograms. They found that these low-latitude plumes are completely analogous to polar plumes. However, such low-latitude plumes have received little attention from the community, partly because they are difficult to discern from the prominent emission in the foreground and background structures \citep{Wang2008} relative to those at higher latitudes. 

Here we report new observational results made possible by the Atmospheric Imaging Assembly \citep{Boerner2010} onboard the Solar Dynamics Observatory. The AIA observations clearly show that upward propagating disturbances (likely outflows) on plume-like structures are not only present at the limb above polar coronal holes, but also in quiet-Sun regions and equatorial coronal holes. We demonstrate that outflows from the quiet Sun may largely contribute to the blueshift of coronal emission lines in adjacent coronal holes. In what follows we explore the implications of these new observational results. 

\section{Observations and results}
The AIA instrument takes full-disk images of the Sun in seven EUV channels and three UV-visible channels. The high-resolution data acquired in some EUV channels, especially the 171\AA{} passband, has an unprecedented signal to noise ratio and thus is excellent for studying faint coronal emission. Inspection of the AIA 171\AA{} data suggests that plume-like structures are commonly present in different regions of the Sun, e.g., polar coronal holes, low-latitude coronal holes, active regions, and quiet-Sun regions. Plume-like structures at active region boundaries, which are usually referred to as fan-like structures, have been studied intensively by using data from  {\it TRACE}  and  {\it Hinode} \citep[e.g.,][]{Berghmans1999,DeMoortel2000,DeMoortel2002,Robbrecht2001,Marsh2003,King2003,McEwan2006,Sakao2007,McIntosh2009a,Marsh2009,Wang2009,He2010a,Tian2011a,Tian2011b,Stenborg2011}, and thus are not discussed here. Here we concentrate on the dynamics of plume-like structures clearly visible in coronal holes and quiet-sun regions.

\subsection{AIA observations of outflows in plume-like structures}

\begin{figure*}
\centering {\includegraphics[width=0.98\textwidth]{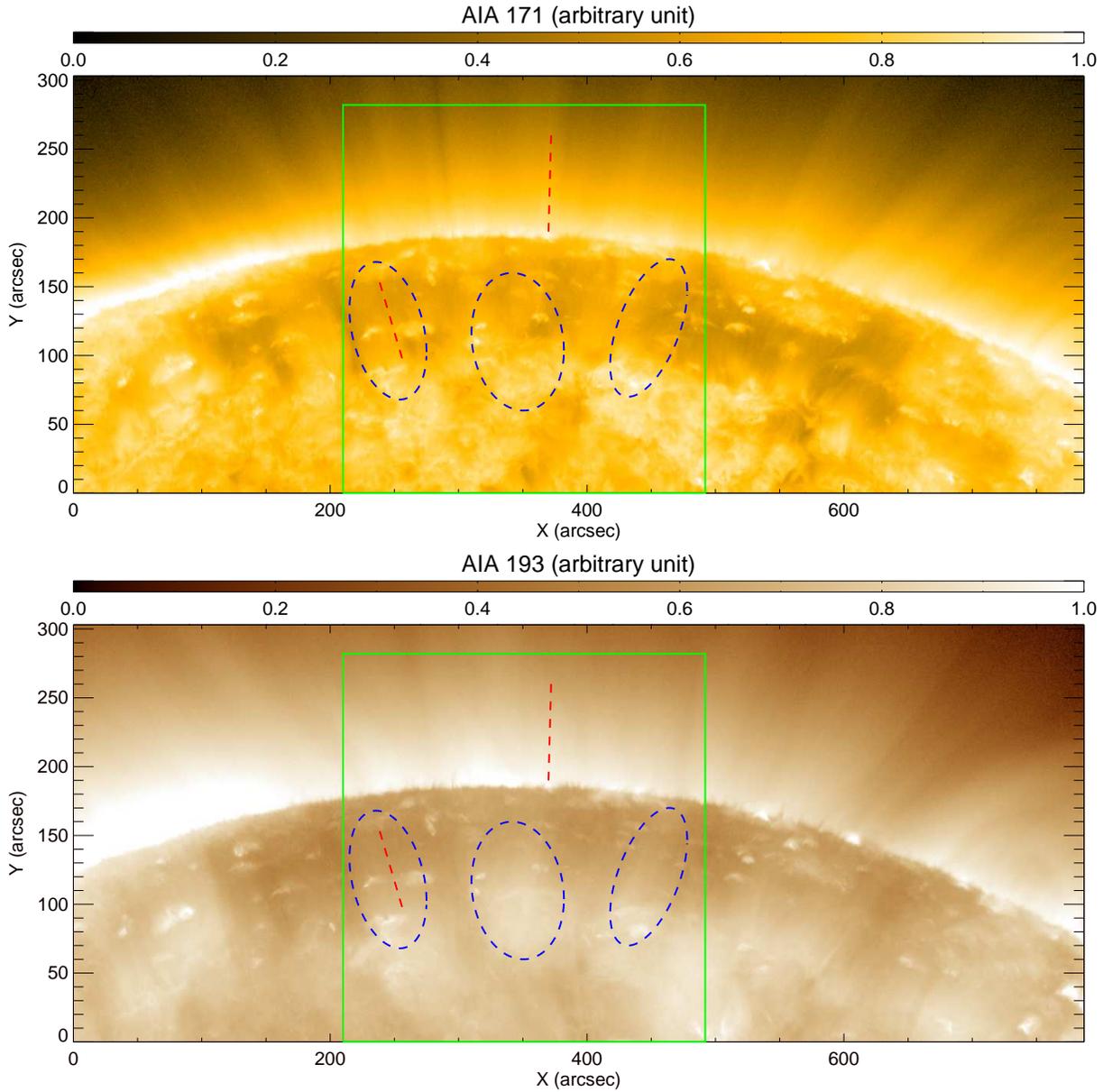}}
\caption{AIA 171\AA{}  and AIA 193\AA{} images of a polar region taken at 22:57 UT on 2010 August 5. Both PCH plumes and QS plumes are present. The ellipses mark the approximate locations of three examples of QS plumes. The two red dashed lines in each panel mark two virtual slits along which the space-time plots in Figure~\ref{fig.4}(A)\&(B) are produced. A movie of the green rectangular region is available online (m1.mpeg).}
\label{fig.1}
\end{figure*}

Figure~\ref{fig.1} shows both AIA 171\AA{} and 193\AA{} images of a polar region, including both the polar coronal hole and the surrounding lower-latitude quiet-Sun region. An online movie consisting of images taken from 22:57 to 23:45 UT on 2010 August 5 shows the evolution of the emission in the outlined rectangular region. To avoid smoothing out the faint plume emission, no de-rotation of the AIA images was performed. The movie reveals that plume-like emission structures are not only present at the limb above the polar hole, but are also clearly visible in the quiet-Sun foreground region. Continuous outward motions are strikingly visible in plumes originating from both the quiet Sun and coronal holes. These outward propagating disturbances are similar in appearance, speed, quasi-periodicity, intensity change, and multi-thermal nature, to the outflows identified by \cite{McIntosh2010} in polar plumes and quasi-periodic rapid upflows inferred from spectral line profiles \cite[e.g.,][]{DePontieu2009,DePontieu2010,Tian2011a}, and thus are probably dominated by outflows rather than compressional waves (see below). A survey of the AIA data suggests that quiet-Sun plumes (QS plumes) are not easy to identify due to the strong emission from foreground and background structure, and that they are easily discernible when projected onto the plane of the sky above the surrounding low-emission coronal holes. Similar to plumes in polar coronal holes (PCH plumes), QS plumes are often associated with coronal bright points rooted in small magnetic bipoles at the photospheric base. We also note that both QS and polar plumes consist of obvious fine structures, and that these structures show non-stop swaying motions indicative of the presence of ubiquitous Alfv\'en waves \citep{Tomczyk2007,DePontieu2007} that will be the subject of a subsequent paper.

\begin{figure*}
\centering {\includegraphics[width=0.98\textwidth]{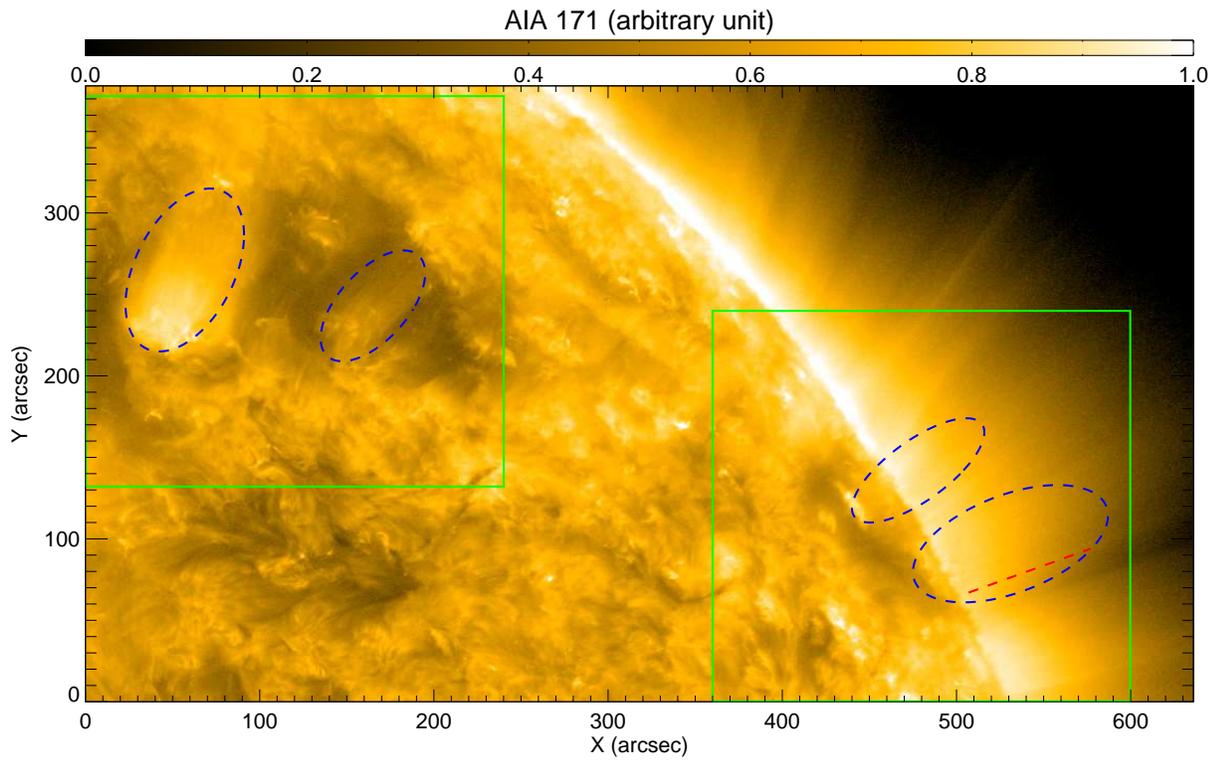}}
\caption{AIA 171\AA{} image of a region around the west limb taken at 00:01 UT on 2010 August 30. The two rectangles enclose two low-latitude coronal-hole regions, respectively. The ellipses mark the approximate locations of four examples of low-latitude plumes. The red dashed line marks a virtual slit along which the space-time plots in Figure~\ref{fig.4}(C) are produced. A movie associated with this figure is available online (m2.mpeg). }
\label{fig.2}
\end{figure*}

Another example of QS plumes can be found in the outlined region at the upper left of Figure~\ref{fig.2}. The associated online movie shows the evolution of the 171\AA{} emission from 00:01 to 01:08 UT on 2010 August 30 in this region. Two well-defined QS plumes, with the eastern one showing much stronger emission, are clearly present at the boundary between the low-latitude (equatorial) coronal hole (ECH) and the surrounding quiet-Sun region. Again, continuous outflows are found in these two plumes. The outlined region in the lower right part of Figure~\ref{fig.2} shows a low-latitude coronal hole at the western limb. The extended structures beyond the western limb (ECH plumes) show a morphology similar to polar plumes. The prominent outward motions and ubiquitous swaying motions associated with these low-latitude plumes also reveal no obvious difference from those in polar plumes. 

\begin{figure*}
\centering {\includegraphics[width=0.98\textwidth]{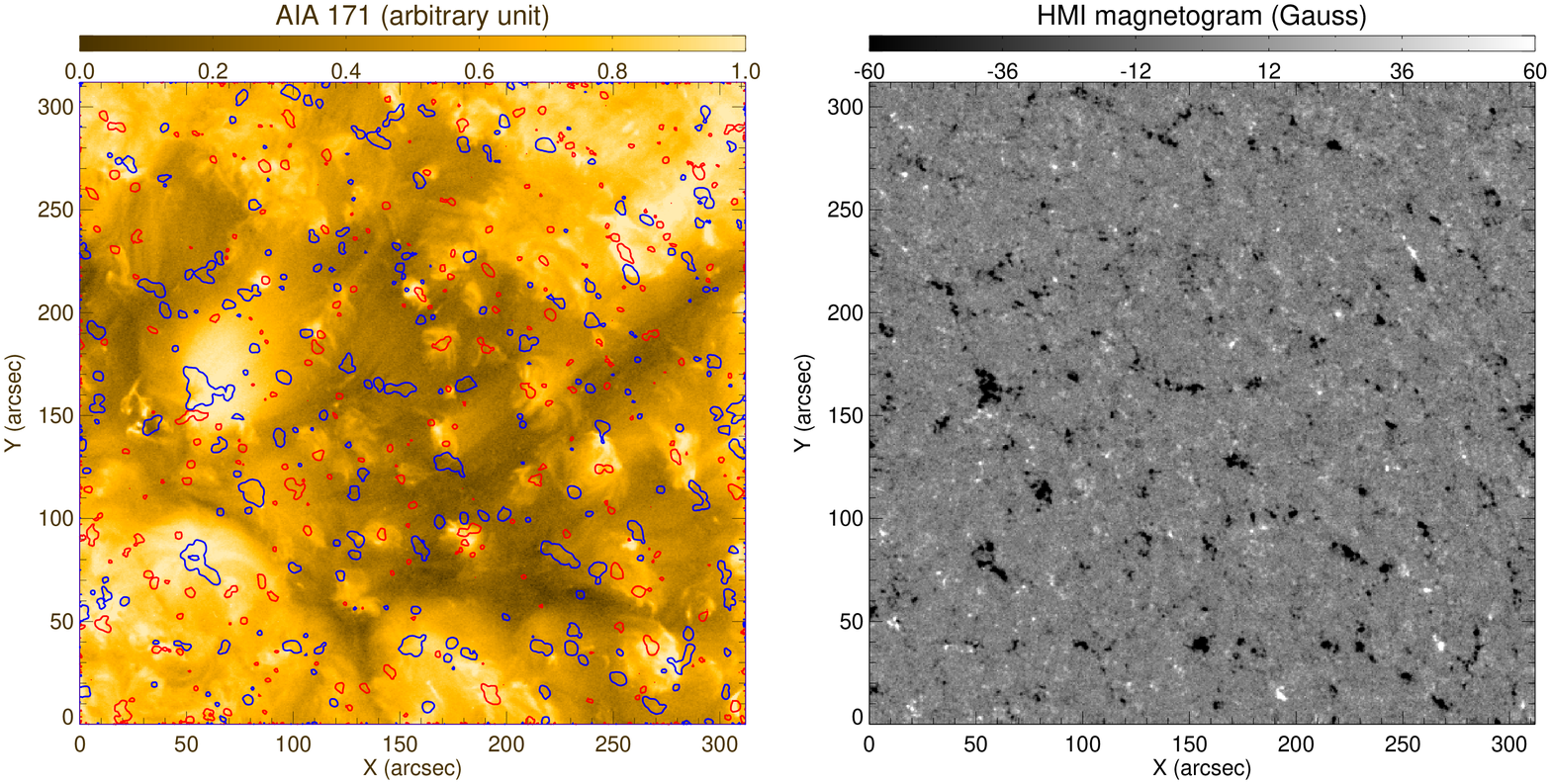}}
\caption{AIA 171\AA{} image and HMI magnetogram of a low-latitude coronal hole located on the northwest quadrant of the solar disk taken at 23:01 UT on 2010 August 25. Red (positive) and blue (negative) contours representing regions of magnetic field strength larger than 15 G are superimposed on the AIA image. ECH plumes are clearly visible inside the coronal hole. A movie associated with this figure is available online (m3.mpeg). }
\label{fig.3}
\end{figure*}

The left panel of Figure~\ref{fig.3} shows a snapshot of a low-latitude coronal hole on the northern hemisphere on 2010 August 25, in the 171\AA{} passband. The associated online movie, which consists of images taken from 23:01 to 23:59 UT, clearly shows the presence of fast outflows along plume-like structures. We also present in the right panel the corresponding magnetogram obtained by the Helioseismic and Magnetic Imager (HMI) onboard SDO. Detailed investigations of the relationship between temporal evolution of the magnetic field and coronal outflows are beyond the scope of this paper and will be performed in the future. From Figure~\ref{fig.3} we can see that low-latitude plumes often originate from strong-field network regions. Mix-polarity magnetic fields are often found around the roots of plumes. Although big plumes seem to be rooted in coronal bright points, we also found some outflows which are not associated with obvious bright points.

\subsection{Speed and period of the outflows}

\begin{figure}
\centering {\includegraphics[width=0.48\textwidth]{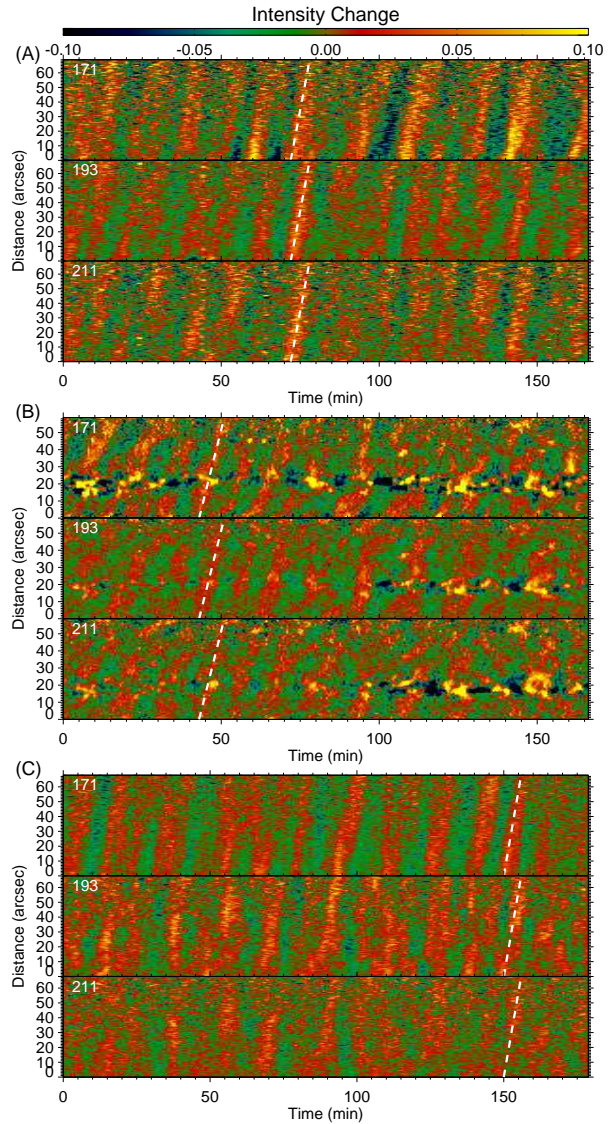}}
\caption{Space-time plots of the detrended intensities in three AIA passbands, for a plume originating from a polar coronal hole (A), a quiet-Sun region (B), and an equatorial coronal hole (C). Locations of the virtual slits are shown in Figure~\ref{fig.1}\&\ref{fig.2}. The starting time is 23:58 UT on 2010 August 05 in (A)\&(B) and 00:01 UT on 2010 August 29 in (C). The inclined dashed line in each panel indicates a well-identified outflow event and its slope represents the projected speed of the outflow. }
\label{fig.4}
\end{figure}

The speed of these outflows can be estimated by placing a virtual slit along the propagation direction of the flow and calculating the slope of a bright strip in the space-time (S-T) plot \citep[e.g.,][]{Sakao2007,McIntosh2010}. To resolve flows associated with different fine structures within one plume, the slit width should not be too large. It can not be too small since the fine structures often show non-stop swaying motions so that propagating paths of the outflows are often not straight. Considering these two effects, the slit width was chosen to be 20 pixels (12$^{\prime\prime}$). As an example, Figure~\ref{fig.4} presents S-T plots for a PCH plume, a QS plume, and an ECH plume. To better reveal the faint outflow signatures here we show the detrended intensities, which were obtained by first subtracting a 15-minute running average from the original intensity time series and then normalized to the running average at each location of the slit. The inclined bright and dark strips indicating propagating intensity perturbations are clearly revealed in the figure. Three examples of well-identified outflow events are marked by the inclined dashed lines in Figure~\ref{fig.4}. The projected speeds of these three outflows were calculated to be 138, 89, and 117 km s$^{-1}$ respectively. 

\begin{figure}
\centering {\includegraphics[width=0.48\textwidth]{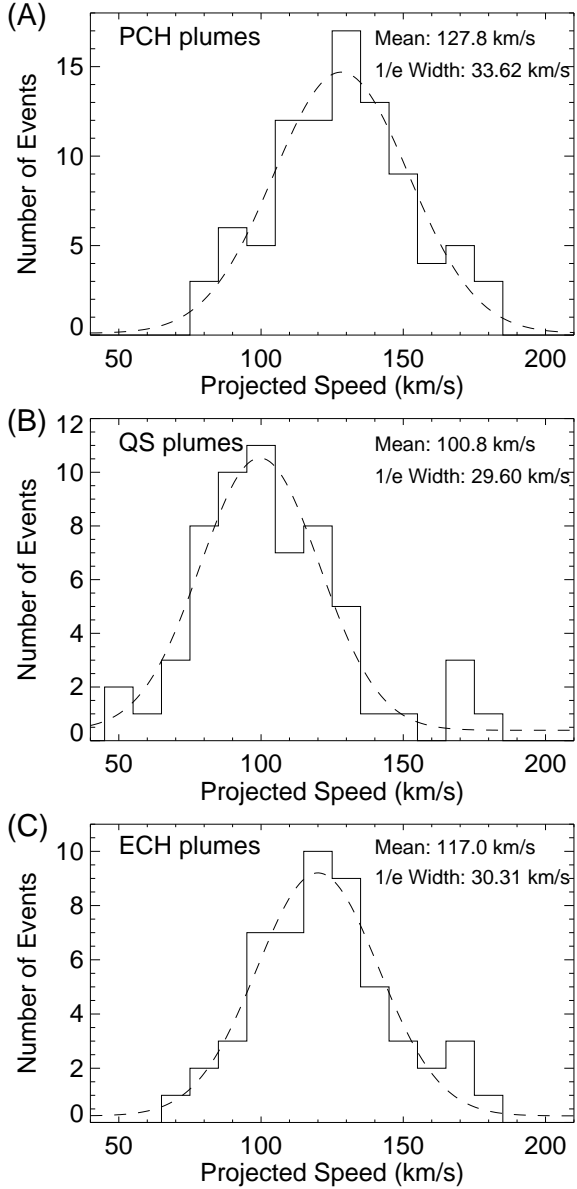}}
\caption{Histograms showing the speed distributions of outflows in a polar coronal hole (A), a quiet-Sun region (B), and an equatorial coronal hole (C). Dashed lines are Gaussian fits to the histograms. The mean value of the speed and the Gaussian width of the fit are shown in each panel. }
\label{fig.5}
\end{figure}

We produced S-T plots for all plume outflows clearly visible in our data and selected well-identified events which reveal clear strips in all the three passbands (171 \AA{}, 193 \AA{}, 211 \AA{}) for statistical analysis. The speed was calculated from the slope of the bright strip for each event. The AIA data obtained from 22:57 UT on 2010 August 5 to 02:43 UT on the next day was used to to produce S-T plots of PCH and QS plumes (see Figure~\ref{fig.1}). For ECH plumes we used the AIA data obtained from 00:00 UT to 03:00 UT on 2010 August 29 and 00:00 UT to 04:00 UT on the next day (see Figure~\ref{fig.2}). In total we have identified 89, 61, and 53 outflow events in the polar coronal hole, quiet Sun, and equatorial coronal hole regions, respectively. We present in Figure~\ref{fig.5} the distributions of the projected speeds of outflows in different regions. A Gaussian fit was applied to each speed distribution and the Gaussian width is shown in each panel. We found that the speed is mainly in the range of 70-180 km s$^{-1}$ and the Gaussian withs of the speed distributions are all about 30 km s$^{-1}$ in different regions. There is no obvious difference of the outflow speed in polar and equatorial coronal holes. Note that the measurements for PCH plumes and ECH plumes were made at the limb, while the roots of QS plumes are located about 100$^{\prime\prime}$ away from the limb. Taking this line of sight effect into account and assuming that the QS plumes are radially aligned on average, the mean speed of flows in these QS plumes is estimated to be 100.8$\times$(960./860.)$\approx$113 km s$^{-1}$, which is not so different from those in PCH plumes and ECH plumes.

These outflows often occur quasi-periodically with a period of 5 to 15 minutes and reveal an intensity change of a few percent. The quasi-periodicity can be clearly seen in the S-T plots. We note that intensity oscillations with similar periods between 1.5 and 2.2 solar radii have been reported, by using data obtained by the Ultraviolet Coronagraph Spectrometer \citep{Morgan2004}. These oscillations might be signatures of recurring high-speed jets similar to what we report here. The patterns of S-T plots are often similar in the three AIA passbands, indicating a similar flow speed and period at different temperatures ranging from 7$\times$10$^{5}$~K to 2$\times$10$^{6}$~K. In most cases we did not find an obvious increasing trend in the speed with increasing temperature,
which seems to support the interpretation of the propagating disturbances as multi-thermal high-speed upflows rather than slow mode waves, since the slow wave speed is expected to be temperature dependent. Such results are consistent with the polar plume study of \cite{McIntosh2010}. However, a recent investigation
suggests that the contribution of the dominant ions (Fe~{\sc{xii}} in the 193 \AA{} passband and Fe~{\sc{xiv}} in the 211 \AA{} passband) to the total emission is usually less than 50\% in the 193 \AA{} and 211 \AA{} passbands \citep{ODwyer2010}, suggesting that the emission of the propagating disturbances in these two passbands might be contaminated significantly by cool ions. Thus, we can not rule out the slow wave interpretation. In fact high-speed upflows can lead to the formation of shocks and trigger slow waves when propagating upward. So it is also likely that both slow waves and upflows are existing in our observations. However, considering the similar speed, quasi-periodicity, and intensity change between these outward propagating disturbances and the quasi-periodic rapid upflows inferred from spectral line profiles, we think that the propagating disturbances are more likely to be, or dominated by, high-speed outflows. Such a conclusion is also supported by the predominant blue shift associated with the propagating disturbances (see Figure~\ref{fig.7}). We note that for some plumes both the emission morphologies and S-T plots show some differences in different AIA passbands, especially between the 171 \AA{} passband and the other two passbands. The emission in the 171 \AA{} passband is dominated by Fe~{\sc{ix}}~171.107\AA{} \citep{ODwyer2010}, which is formed in the upper transition region and thus may have a smaller scale height compared to the hotter emission in the other two passbands. Moreover, due to the different sensitivity, signal to noise ratio, and emission contrast in different passbands we may not always be able to clearly identify every outflow event in all the three passbands from the S-T plots. Note that the 211 \AA{} passband has the lowest signal to noise so that outflow events are sometimes revealed not as clearly as in the other two passbands. 

\subsection{Impact of quiet-Sun outflows on spectroscopic observations of the adjacent coronal holes}

\begin{figure*}
\centering 
\begin{minipage}[t]{0.48\textwidth}
{\includegraphics[width=\textwidth]{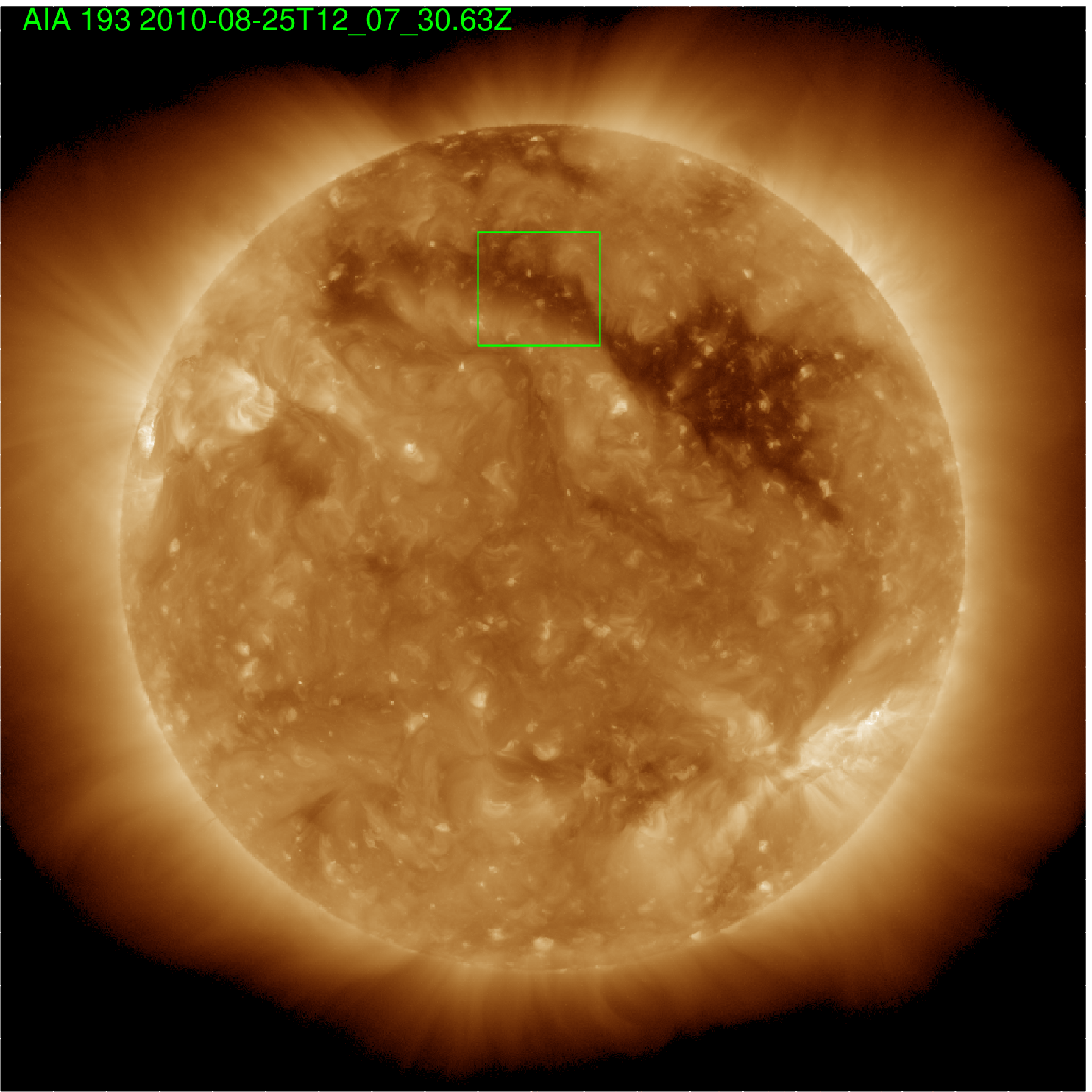}}
\end{minipage}
\begin{minipage}[t]{0.48\textwidth}
{\includegraphics[width=\textwidth]{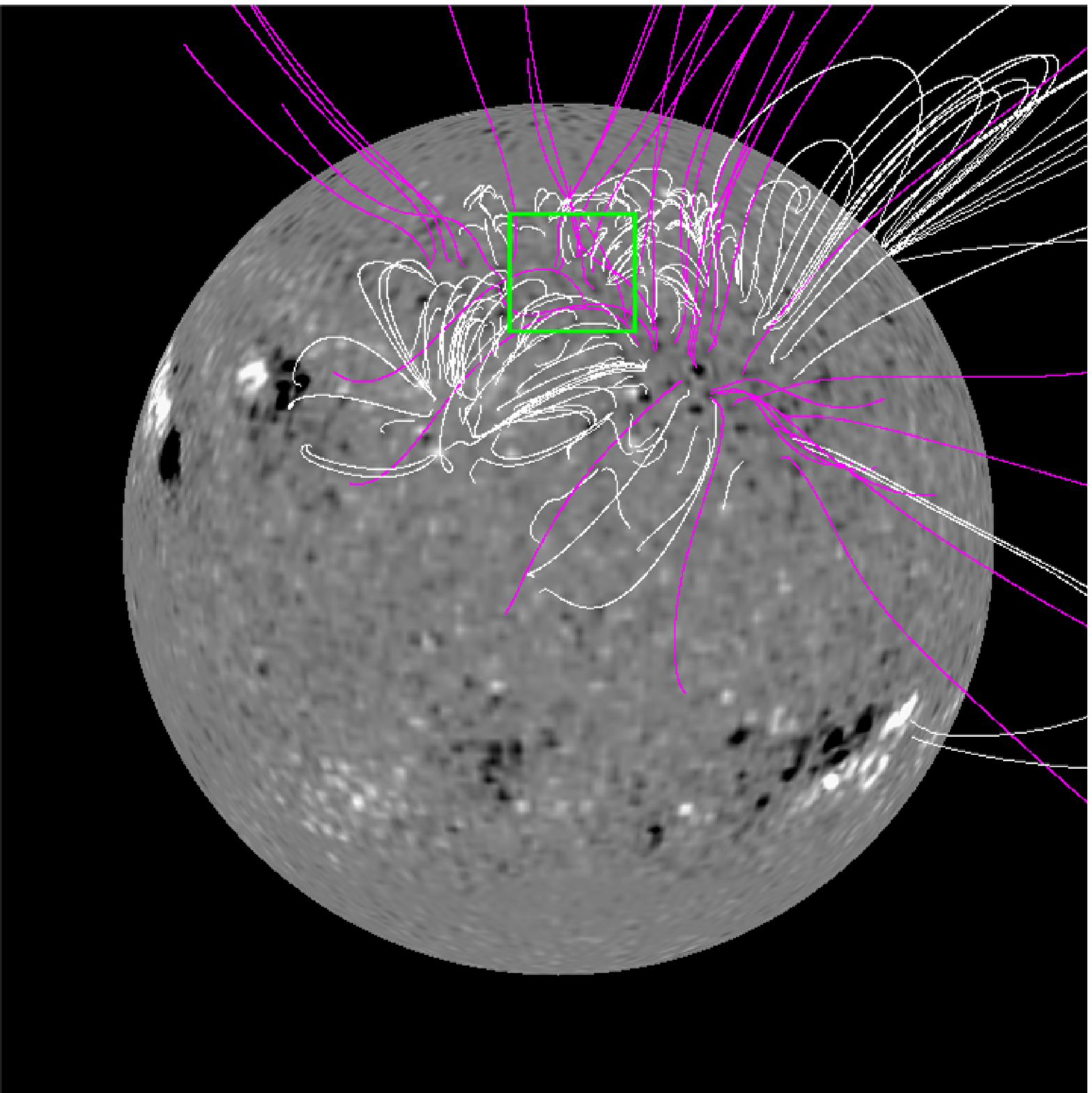}}
\end{minipage}
\begin{minipage}[b]{\textwidth}
\caption{ Left: Full disk image of AIA 193\AA{} at 12:07 UT on 2010 August 25. Right: Result of PFSS model at 12:06 UT on 2010 August 25. White and purple lines represent closed and open magnetic field lines, respectively. The green rectangle in each panel outlines the field of view shown in Figure~\ref{fig.7}. }
\label{fig.6}
\end{minipage}
\end{figure*}

\begin{figure*}
\centering {\includegraphics[width=0.98\textwidth]{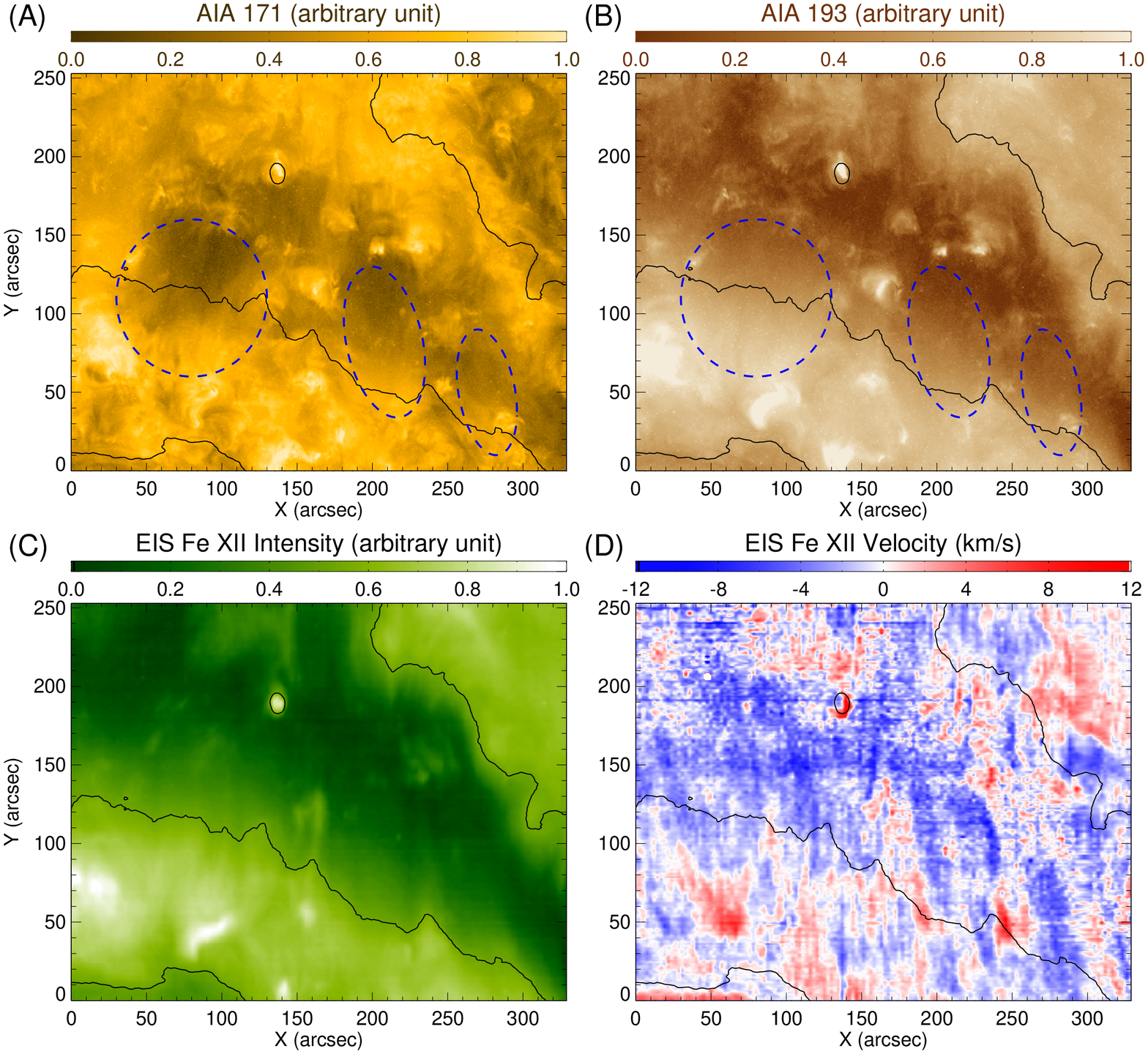}}
\caption{(A)-(B) AIA 171\AA{} and 193\AA{} images of a low-latitude coronal hole taken at 10:14 UT on 2010 August 25. (C)-(D) Intensity map and Dopplergram of EIS Fe~{\sc{xii}}~195.12\AA{} obtained through a single Gaussian fit. The EIS scan was performed from 10:14 to 14:56 UT. The black lines represent the coronal hole boundary. Examples of QS plume outflows contaminating the blueshift patches on Dopplergrams are marked by ellipses in (A) and (B).  Movies associated with (A) and (B) are available online (m71.mpeg, m72.mpeg).}
\label{fig.7}
\end{figure*}

From Figures~\ref{fig.1}\&\ref{fig.2} and the associated movies we can clearly see that quiet-Sun plumes often project onto the plane of the sky above coronal holes. This finding has direct impact on the interpretation of blue-shifted patches present in Dopplergrams of coronal emission lines in coronal holes. These blue shifts have been widely interpreted as the nascent fast solar wind from coronal holes \citep[e.g.,][]{Hassler1999,Peter1999,Stucki2000,Wilhelm2000,Xia2003,Tu2005,Aiouaz2005,Tian2010a}. However, a coordinated observation of AIA and the EUV Imaging Spectrometer \citep[EIS,][]{Culhane2007} onboard Hinode as shown in Figure~\ref{fig.7} and the associated movies (m71.mpg for AIA 171\AA{} and m72.mpg for 193\AA{}) reveal clearly that the blueshifts of Fe~{\sc{xii}}~195.12\AA{} in the low-latitude coronal hole are largely contaminated by outflows from the surrounding quiet Sun. In Figure~\ref{fig.6} we show a full-disk image of AIA 193\AA{}, and the magnetic field structures obtained by using the potential field source surface (PFSS) model \citep{Schrijver2003} in and around the targeted coronal hole. In Figure~\ref{fig.7} the black lines which are determined by using a threshold of the Fe~{\sc{xii}}~195.12\AA{} intensity outline the boundary of the coronal hole. Through a comparison between the AIA movies and the Fe~{\sc{xii}}~195.12\AA{} intensity images and Dopplergrams we can conclude that the contribution of outflows from the quiet Sun dominates the emission of most blue-shifted patches in the coronal hole. This is because the quiet-Sun plumes projected above the coronal hole are denser than the background coronal hole plasma and significantly contribute to the emission detected by the spectrograph. A direct comparison of the flow speed derived through imaging observations with that from spectroscopic observations is not possible since the orientation of the plumes can not be precisely determined without stereoscopic observations. Note that the AIA 171\AA{} data has a much better signal to noise ratio than the data acquired in the 193\AA{} passband which samples plasma at a temperature much closer to the formation temperature of Fe~{\sc{xii}}~195.12\AA{}. 

\begin{figure*}
\centering {\includegraphics[width=0.98\textwidth]{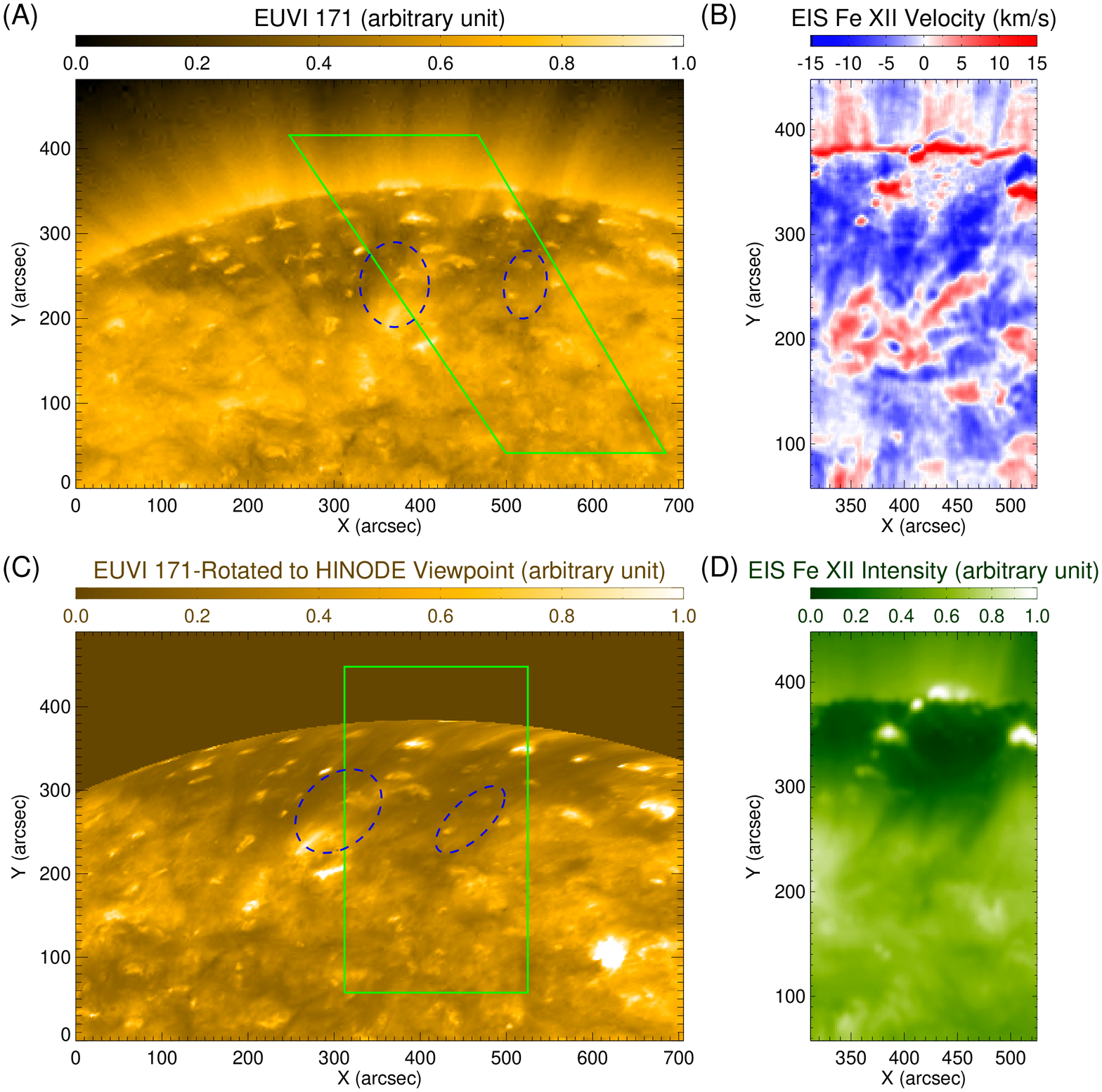}}
\caption{(A) EUVI/SECCHI 171\AA{} image of a polar coronal hole and the surrounding quiet-Sun region taken by STEREO-B at 14:14 UT on 2007 October 10. (C) The EUVI/SECCHI 171\AA{} image rotated to the viewpoint of HINODE. The regions outlined in green mark the approximate location where the EIS observation was made. (B)\&(D) Dopplergram and intensity map of the EIS Fe~{\sc{xii}}~195.12\AA{} line obtained through a single Gaussian fit. The EIS scan was performed from 14:13 to 18:17 UT. Examples of QS plume outflows contaminating the blueshift patches on Dopplergrams are marked by ellipses in (A) and (C). A movie associated with (A) is available online (m8.mpeg). }
\label{fig.8}
\end{figure*}

Figure~\ref{fig.8} provides another example of the complexity of blueshifts in a polar coronal hole. This EIS data set was taken on 2007 October 10, when the STEREO-B spacecraft was 17 degrees different from the Earth in Heliocentric Inertial (HCI) longitude. The EIS data was previously used by \cite{Tian2010a} to study coronal hole outflows. In Figure~\ref{fig.8}(A)\&(C) we show an original image of EUVI 171\AA{} and the result after rotating it to the viewing point from the HINODE spacecraft. The green outlined regions show the approximate location where the EIS scan was performed. We did not do any rotation of the EUVI images when making the movie since it would smooth out the faint plume emission and outflow signatures, as demonstrated by Figure~\ref{fig.8}(C). The signatures of outflows in QS plumes are much weaker but still visible in the EUVI movie as compared to the AIA movies shown above. It is clear that some blue-shift patches, especially those close to the coronal hole boundary (see the locations marked by ellipses), are contaminated by outflows in QS plumes. The contamination should be more significant with increasing temperature, since the coronal plasma is hotter in the quiet Sun than in coronal holes. Assuming that QS plumes and PCH plumes have a similar density, and that the background coronal hole density is comparable to that of the inter-plume regions above the limb, we estimate the contribution from QS plumes to the emission of these blue-shift patches to be around 57\% at log({\it T}/K)=6.0 and 72\% at log({\it T}/K)=6.2 if the QS plumes are not superimposed on PCH plumes. Thus, caution must be taken when interpreting the blueshifts of coronal emission lines derived from spectroscopic observations, and particularly those without sufficient S/N and spectral resolution to decompose the various components of the emission.

If the QS plumes are radially aligned, the contamination should be more significant at higher latitudes since the projection effect is more significant there. However, offlimb observations reveal that the plume orientation can sometimes deviate largely from the local radial direction. So the contamination by QS emission may not be negligible even in some low-latitude CHs. Due to the contamination of QS plasma, not only the study of the Doppler shift, but also some other investigations such as density and temperature diagnostics in coronal holes might need to be reconsidered. However, since most of the past imaging observations had signal to noise ratios too poor to reveal the faint QS plumes, we are not able to quantitatively evaluate the importance of the contamination effect in these previous studies. We note that the importance of the projection effect has also been discussed in the context of polar rays \citep{Li2000} and magnetic loops as well as coronal jets \citep{He2010b}. 

\section{Discussion}
\subsection{Implications for coronal heating}
The AIA observations presented here reveal almost ubiquitous outward propagating disturbances in the quiet Sun and coronal holes. These outward motions are distinctly visible in plume-like structures. They are quasi-periodic (with a period of 5 to 15 minutes), high-speed ($\sim$120$\pm$30 km s$^{-1}$), often multi-thermal (at least in the temperature range of 7$\times$10$^{5}$~K to 2$\times$10$^{6}$~K), and show an intensity modulation of a few percent. These observational results are similar to those of the propagating disturbances observed along the fan-like structures at the edges of active regions \citep[e.g.,][]{Berghmans1999,DeMoortel2000,DeMoortel2002,Robbrecht2001,Marsh2003,King2003,McEwan2006,Sakao2007,McIntosh2009a,Marsh2009,He2010a,Tian2011a,Tian2011b,Stenborg2011}. With the unprecedented high-quality AIA observations, we can now conclude that high-speed quasi-periodic propagating disturbances are not restricted to polar plumes and AR fans, but clearly present almost everywhere in the corona. The fact that they are more easily observed in certain regions is probably related to the simple magnetic structures and weak background/foreground emission there (e.g., polar plumes, AR edges).   

Through analyses of the asymmetries of EUV spectral line profiles, \cite{DePontieu2009} and \cite{McIntosh2009b} identified ubiquitous faint upflows (exhibit as weak blueward excess emission) with a lifetime of 50-150 seconds and a speed of 50-150 km s$^{-1}$  in magnetized regions of the quiet Sun, coronal holes, and active regions. These faint upflows are believed to be associated with type-II spicules or rapid blue-shifted events observed in the chromosphere \citep{DePontieu2009,Rouppe2009}. They are suggested to provide hot plasmas into the corona and may thus play an important role in coronal heating process \citep {DePontieu2009,McIntosh2009b,Peter2010,DePontieu2010,Hansteen2010,DePontieu2011}. 

The period, speed, multi-thermal nature, and intensity fluctuation of the coronal disturbances reported here are all similar to those of the quasi-periodic upflows inferred from blue-wing asymmetries of EUV spectral line profiles \citep{DePontieu2009,McIntosh2009a,McIntosh2009b,McIntosh2011,DePontieu2010,Tian2011a,Tian2011b,Martnez-Sykora2011,Ugarte-Urra2011}, indicating that they are likely to be the same phenomenon and that the coronal disturbances are probably dominated by upflows rather than slow-mode waves \citep{McIntosh2010,DePontieu2010,Tian2011a,Tian2011b}. The flow interpretation of the coronal disturbances is also supported by the same speed in different AIA passbands and the association of the disturbances with the blue shift of EIS emission lines, as mentioned above. In edges of ARs, observations show clearly that the outward motions along fan-like structures are responsible for the blue-wing asymmetries of coronal line profiles and that they may provide heated mass into the corona \citep{McIntosh2009a,Tian2011a,Tian2011b,DePontieu2011}. Unfortunately, the EIS spectra in the quiet Sun and coronal holes are usually too weak to allow a reliable analysis of the asymmetries of line profiles, making it difficult to establish a direct connection of the weak blueward excess emission and outward motions (plume outflows) in these regions. However, based on the similarities mentioned above, we conjecture that the QS and CH outflows we report here might be an important means to provide heated mass into the corona and solar wind. 

\subsection{Difference between plumes and inter-plume regions?} 
Observations seem to suggest that polar plumes only occupy a small portion of polar regions \citep[e.g.,][]{Wilhelm2006,Curdt2008} and often reveal a smaller spectral line width and lower flow speed compared to the inter-plume regions, leading to the conclusion that plumes are not the main source regions of the fast solar wind \citep[e.g.,][]{Wang1994a,Wilhelm1998,Giordano2000,Patsourakos2000,Banerjee2000,Wilhelm2000,Teriaca2003,Raouafi2007,Feng2009}. However, other studies suggest that the flow speed in plumes can be higher than those of the inter-plume regions and that plumes are a substantial contributor of mass to the fast solar wind \citep{Casalbuoni1999,Gabriel2003,Gabriel2005,McIntosh2010}. Further, in studying the compositional difference between plume and inter-plume regions \cite{DelZanna2003} found no obvious FIP effect (enrichment of low first-ionization potential elements) in plumes, also questioning the argument that plumes are not the primary source of material of the fast solar wind. 

Closer inspections of our on-line movies reveal that outflows of similar speeds are also present in inter-plume regions. The inter-plume outflows are usually not obviously associated with coronal bright points. The discoveries of plume-like structures in the quiet Sun and equatorial coronal holes, and the continuous high-speed outflows from both plume and inter-plume regions may challenge the current knowledge of plumes and solar wind origin. As mentioned above,  previous debate concentrated on whether the solar wind outflow originates from plumes or inter-plume regions in polar coronal holes. Our findings reveal that both of them are tunnels of high-speed outflows, part of which may efficiently feed the solar wind. It appears that plumes in both coronal holes and quiet-Sun regions are nothing more than locally denser regions of plasma - this enhanced density ensures that the outflows are more easily visible. We stress that the smaller line widths (if regarded as a proxy of wave amplitude) in plumes compared to inter-plume regions might just be an effect of density, given that the input Alfv\'en wave flux is similar everywhere \citep{Doschek2001}. In fact, our movies reveal that the boundaries between plumes and inter-plumes are not always clearly defined, and that the swaying motions of plumes can change the boundaries substantially, which may lead to an impression of mass transfer from plumes to inter-plume regions \citep{Gabriel2005}.  

\subsection{Solar wind origin from the quiet Sun?}
The high-speed outflows in our AIA observations usually reach rather high (often clearly visible to heights $\sim$100~Mm above their sources) and thus they may be channeled along large coronal loops or open field lines. There is little doubt that the plume flows are guided by open field lines in both polar and equatorial coronal holes. Some of these coronal hole outflows may overcome the gravity, thus being an important source of mass supply to the solar wind. It seems that the QS plumes we discovered are likely to be the lower parts of large coronal loops, since the quiet Sun is characterized by closed fields according to the traditional view of solar magnetic field structures. However, based on coronagraph observations, \cite{Habbal2001a} suggested that the coronal magnetic field is predominantly radial, and that the solar magnetic field consists of two components: a dipole-like field associated with large-scale structures and a radial field associated with the pervasive open field lines originating from both the quiet Sun and coronal holes. Coronagraph observations reveal that the latitudinal density profiles at different heights are similar, indicating that polar coronal holes extend radially into interplanetary space \citep[e.g.,][]{Woo1997,Woo2007}. \cite{Luo2008} also concluded that high-latitude coronal holes have little influence on fast wind streams at 1 AU. The quiet Sun as a possible source of the solar wind has been proposed based on investigations of the latitude dependence of solar wind velocity \citep{Habbal1997,Habbal2001,Woo2000} and correlation analysis of EUV observations with magnetic fields \citep{He2007,Tian2008,Tian2010b}.

The right panel of Figure~\ref{fig.6} shows the coronal magnetic field structures calculated from the PFSS model in and around a coronal hole. A comparison of this magnetic field configuration with the QS outflow trajectories revealed in the associated movies (m71.mpg, m72.mpg) seems to suggest no good agreement between the loop legs and flow paths: the flow paths are apparently much more inclined towards the pole compared to the loop legs in the lower part of the selected region. This inconsistency may suggest that the traditional dipole-like PFSS is not valid to explain the QS outflow trajectories or that there might be a second component of the solar magnetic field besides the traditional dipole-like field. If the latter is the case and the second component is open \citep{Habbal2001a}, our observation of plume outflows with similar properties in both the quiet Sun and coronal holes is probably providing support to the scenario of solar wind origin from both regions. The possible presence of open flux in closed-field regions is also implied by the interchange reconnection model, which predicts that open flux can diffuse deep inside closed-field regions \citep{Fisk1998,Fisk2003,Fisk2009}. However, recent theoretical investigations indicate that it is hard to bring open fields into closed-field regions without having them close down \citep {Antiochos2007,Antiochos2011,Titov2011,Linker2011}. Thus, we can not exclude the possibility of some QS plumes being the legs of large coronal loops. We realize that more detailed work, especially stereoscopic observations and 3-D reconstructions \citep[e.g.,][]{Feng2007,Aschwanden2008,Feng2009}, are needed to study the magnetic field structures associated with QS plumes and whether these outflows can be traced outward to the interplanetary space.

\subsection{Magnetic reconnection at coronal hole boundaries?}

We found a paper submitted to arXiv.org by \cite{Yang2011} who reported repetitive EUV jets occurred at coronal hole boundaries by using AIA 193\AA{} data when our paper was under review. These authors focused on a coronal hole boundary at middle latitudes and found some jets associated with the emergence and cancelation of magnetic fields. They claimed that these jets are signatures of magnetic reconnection and that they maintain the rigid rotation of coronal holes \citep[e.g.,][]{Fisk1999,Wang1994}. We think that these jets are similar to the high-speed outflows associated with the fine strands in plume-like structures in our observations. We should also point out that these jets are not only restricted to coronal hole boundaries, but also exist inside coronal holes, quiet-Sun regions, and active regions. The outflow speeds we report here are of the same order compared to the speeds of EUV and X-ray jets \citep[e.g.,][]{Shibata1992,Shimojo1996,Savcheva2007,Cirtain2007,Kamio2010}. Since it is generally accepted that magnetic reconnection might be the triggering mechanism of EUV and X-ray jets \citep[e.g.,][]{Shibata1994,Cirtain2007,Subramanian2010,Edmondson2010a}, we propose that the high-speed outflows we report here might also be related to magnetic reconnections. Such a conjecture is also supported by our observational result that plume-like structures are often rooted around mixed-polarity field regions (such as Figure~\ref{fig.3}). Numerical simulations have shown that photospheric motions can easily induce rapid current sheet formation and efficient reconnection between small-scale bipoles and open fields \citep[e.g.,][]{Edmondson2010a}. These interchange reconnections are likely to be the production mechanisms of the outflows in open-field-dominant regions (such as Figure~\ref{fig.7}). More detailed analysis of the outflows (or jets) and the underlying magnetic field evolution should be performed in the future. 

\section{Conclusion}
In conclusion, we have found repetitive outflows (jets) at temperatures of million degrees from both the quiet Sun and coronal holes. The outflows are clearly visible in plume-like structures, with an average speed around 120 km s$^{-1}$, a quasi-period of 5-15 minutes, and an intensity modulation of a few percent. Outflows are also visible in the weak-emission inter-plume regions throughout the atmosphere. We have demonstrated that the blueshifts of coronal emission lines observed in coronal holes are not necessarily the signatures of fast solar wind origins from coronal holes, but can be significantly contaminated by outflows from the surrounding quiet-Sun regions.  

\begin{acknowledgements}
{\it SDO} is the first mission of NASA$^{\prime}$s Living With a Star (LWS) Program. EIS is an instrument onboard {\it Hinode}, a Japanese mission developed and launched by ISAS/JAXA, with NAOJ as domestic partner and NASA and STFC (UK) as international partners. It is operated by these agencies in cooperation with ESA and NSC (Norway). The SECCHI/STEREO data used here were produced by an international consortium of the NRL, LMSAL, and NASA GSFC (USA), RAL and Univ. Bham (UK), MPS (Germany), CSL (Belgium), IOTA and IAS (France). Scott W. McIntosh is supported by NASA (NNX08AL22G, NNX08BA99G) and NSF (ATM-0541567, ATM-0925177). Hui Tian is supported by the ASP Postdoctoral Fellowship Program of National Center for Atmospheric Research (NCAR). The National Center for Atmospheric Research is sponsored by the National Science Foundation. Hui Tian thanks B. C. Low, L. Ofman, J. Gosling, B. De Pontieu, C.-Y. Tu, I. De Moortel, J. Mart\'{\i}nez-Sykora, and J. K. Edmondson for helpful discussions. We thank the anonymous referee for his/her efforts to improve our manuscript.
\end{acknowledgements}

\end{document}